\relax
\documentclass[letterpaper]{article} 
\usepackage[hyphens]{url}  
\usepackage{graphicx}  
\usepackage[margin=1in,headheight=25pt]{geometry}
\usepackage{amsmath}
\usepackage{amssymb}
\usepackage{bm}
\usepackage{amsthm}
\usepackage{algorithmicx}
\usepackage{subcaption}
\usepackage{natbib}
\usepackage{booktabs}
\usepackage{enumitem}
\usepackage{fancyhdr}
\theoremstyle{definition}

\usepackage{algpseudocode,algorithm,algorithmicx}

\renewcommand{\cite}{\citep}

\begin{document}
\title{Preliminary Results from a Peer-Led, Social Network Intervention, Augmented by Artificial Intelligence to Prevent HIV among Youth Experiencing Homelessness}
\author{Eric Rice, Ph.D.$^\text{a}$, Laura Onasch-Vera, MSW$^\text{a}$, Graham T. DiGuiseppi, Sc.M$^\text{a}$,\\Bryan Wilder, B.S.$^\text{b}$, Robin Petering, Ph.D.$^\text{c}$, Chyna Hill, Ph.D.$^\text{a}$,\\ Amulya Yadav, Ph.D.$^\text{d}$, Milind Tambe, Ph.D.$^\text{b}$\\\\
\fontsize{10pt}{10pt}\selectfont$^\text{a}$Suzanne Dworak-Peck School of Social Work and Center for Artificial Intelligence in Society,\\ \fontsize{10pt}{10pt}\selectfont University of Southern California\\
\fontsize{10pt}{10pt}\selectfont $^\text{b}$Center for Research on Computation and Society,\\\fontsize{10pt}{10pt}\selectfont John A. Paulson School of Engineering and Applied Sciences, Harvard University\\
\fontsize{10pt}{10pt}\selectfont$^\text{c}$Lens Co\\
\fontsize{10pt}{10pt}\selectfont$^\text{d}$College of Information Sciences and Technology, Pennsylvania State University\\\\
\fontsize{10pt}{10pt}\selectfont Corresponding author: Eric Rice, Ph.D., University of Southern California\\
\fontsize{10pt}{10pt}\selectfont Suzanne Dworak-Peck School of Social Work, 669 W 34th St, Los Angeles, CA. 90089, Email: ericr@usc.edu\\\\
\fontsize{10pt}{10pt}\selectfont This research was supported in part by a grant from the California HIV/AIDS Research Program. 
}
\date{}
\maketitle
\clearpage
\begin{abstract}\thispagestyle{plain}

\noindent \textbf{\textit{Introduction}}: Each year, there are nearly 4 million youth experiencing homelessness (YEH) in the United States with HIV prevalence ranging from 3 to 11.5\%. Peer change agent (PCA) models for HIV prevention have been used successfully in many populations, but there have been notable failures. In recent years, network interventionists have suggested that these failures could be attributed to PCA selection procedures. The change agents themselves who are selected to do the PCA work can often be as important as the messages they convey. To address this concern, we tested a new PCA intervention for YEH, with three arms: (1) an arm using an artificial intelligence (AI) planning algorithm to select PCA, (2) a popularity arm--the standard PCA approach--operationalized as highest degree centrality (DC), and (3) an observation only comparison group (OBS).\\

\noindent \textbf{\textit{Methods}}:  YEH were recruited from three drop-in centers who provide food and crisis management services to YEH in Los Angeles, CA, yielding a sample size of 714. Youth were consented and completed a survey which collected self-reported data on HIV testing, condom use, HIV knowledge, and social network information. Each arm (AI, DC, and OBS) was deployed at each agency with at least 6 months between deployments to allow for new youth to enter the networks at these agencies. For follow ups, 472 youth (66.5\% retention) were interviewed immediately after intervention deployment (1-month post-baseline) and 415 youth (58.5\% retention) were interviewed 3 months post-baseline. Network data was used to select PCAs for the DC arm and for the AI arm. In each intervention arm (AI and DC), 20\% of youth were selected as PCAs, who were then given a 4-hour initial training, followed by 7 weeks of 1 hour follow up sessions. Youth disseminated messages promoting HIV testing and condom use. \\

\noindent\textbf{\textit{Results}}:   Using GEE models, there was a significant change over time ($p<.01$) and a significant time-by-arm interaction ($p<.05$) for condomless anal sex, with AI showing the most rapid decrease in condomless anal sex. The AI arm had rates of condomless anal sex at 24.0\% at baseline, 16.8\% at one month and 14.0\% at three months; relative to DC which reported 38.5\%, 36.9\%, and 29.5\%; and relative to CG which reported 21.4\%, 20.9\%, and 23.5\% respectively.  There was a significant change over time ($p<.01$) for condomless vaginal sex, but no arm interaction. Likewise, there was an increase in new HIV testing across all three arms ($p<.01$), but no arm interaction. There was a significant increase in HIV testing ($p<.01$) and HIV knowledge over time among PCA youth ($p<.01$).\\

\noindent\textbf{\textit{Conclusions}}: PCA models that promote HIV testing, HIV knowledge, and condom use are efficacious for YEH. Both the AI and DC arms showed improvements over time. AI-based PCA selection led to better outcomes and increased the speed of intervention effects. Specifically, the changes in behavior observed in the AI arm occurred by 1 month, but not until 3 months in the DC arm. Given the transient nature of YEH and the high risk for HIV infection, more rapid intervention effects are desirable.\\


\end{abstract}
\clearpage

\section{Introduction}

The most recent estimates suggest that there are 4.2 million adolescents (ages 13-17) and young adults (ages 18-24) who experience homelessness each year in the United States (Morton et al., 2018). These youth who experience homelessness (YEH) are at great risk for contracting HIV/AIDS, with prevalence rates reported as high as 11.5
Given the central role that peers play in the HIV risk and protective behaviors of YEH (e.g., Green, et al., 2013; Rice, 2010; Rice, Barman-Adhikari, et al., 2012; Valente \& Auerswald, 2013), commentators have suggested that a PCA model for HIV prevention should be developed for YEH (Arnold \& Rotheram-Borus, 2009; Rice, Barman-Adhikari, et al., 2012). PCA models identify a small number of individuals in a high-risk target population to become advocates in their community. These individuals are tasked with disseminating HIV prevention information and norm-changing messages to their peers (Amirkhanian, Kelly, Kabakchieva et al., 2003; Kelly, et al., 1997; Latkin, Sherman, \& Knowlton, 2003). 

PCA models have been found to be effective for HIV prevention in many contexts (Kelly et al., 1997; Latkin et al., 2003; Medley, Kennedy, O’Reilly, \& Sweat, 2009), but there have been some notable failures (Kelly, 2004; NIMH Collaborative HIV/STD Prevention Trial Group, 2010). These findings led Schneider and colleagues (2015) to suggest that PCA model failures may be due to how PCAs are selected. They argued that the change agents who are selected to do the PCA work can often be as important, if not more important, than the messages they convey (Schneider et al., 2015). As developed by Kelly, the standard method for selecting PCAs is to use ethnographic methods to identify the most popular individuals in the social network. This can be operationalized more formally as selecting PCAs who have the greatest number of network connections to others in a population, a concept known as highest degree centrality in social network terminology. Several authors—but particularly Valente (e.g., Valente \& Pumpuang, 2007)—have described how network-driven prevention programs can benefit from explicitly modeling social networks and leveraging network methods such as degree centrality in the context of intervention delivery.  

Recent work has focused on the development of AI methods to improve the process of selecting PCAs (Wilder et al., 2017; Wilder, Immorlica, et al., 2018; Wilder, Onasch-Vera, et al., 2018; Yadav et al., 2015; Yadav et al., 2016; Yadav et al., 2018). While AI is still relatively new to many researchers in medicine, public health, and social work, techniques from AI, particularly machine learning approaches, have gained traction in recent years (e.g., Ahn et al., 2016; Char et al., 2018; Verghese et al., 2018). AI, however, is much broader than machine learning and has the potential for varied impact. In general, AI explores the creation of computer software capable of “intelligent behavior,” such as complicated decision-making. Our work was based on influence-maximization research in computer science (Kempe et al., 2003; Yadav et al., 2015; Yadav et al., 2016). This approach involves creating an explicit mathematical model of hypothetical information dissemination in social networks. Then various computational models are developed and tested using observational data from actual networks of YEH. These proscriptive algorithms are then field tested in intervention pilot studies. From a technical standpoint, the AI has been vetted by the peer-review process in computer science and a series of papers has been published showing the evolution of this work (Wilder et al., 2017; Wilder, Immorlica, et al., 2018; Wilder, Onasch-Vera, et al., 2018; Yadav et al., 2015; Yadav et al., 2016; Yadav et al., 2018).

This AI-driven line of work has led to promising preliminary results in pilot studies in the field, suggesting that PCA models can likely be deployed to enhance HIV prevention efforts with YEH in the context of drop-in centers (Rice et al., 2018; Wilder, Onasch-Vera, et al., 2018; Yadav et al., 2017). Drop-in service centers are safe havens where youth can access food, clothing, and case management services. Such programs have been shown to be better suited to engaging YEH in services than emergency shelters (Slesnick et al., 2016). Drop-in centers serve the broadest possible spectrum of YEH, including youth who are living outdoors, youth who stay in emergency shelters, and youth who “couch surf” (or are temporarily doubled-up with friends or family). Developing an effective PCA model that serves these youth from the context of these “low barrier” services settings is a promising, potentially scalable prevention strategy for addressing long term health disparities in HIV (Rice et al., 2018).  

What remains to be seen is whether a PCA intervention for YEH can produce statistically significant improvements in HIV prevention behaviors when powered sufficiently. Thus the purpose of this paper is to present results of a quasi-experimental design, which compares a PCA model delivered in drop-in centers with three study arms: (1) PCA selection based on AI; (2) PCA selection based on popularity, operationalized as YEH with highest degree centrality (DC); and (3) observation only control group (OBS), wherein youth receive usual care with respect to HIV prevention in drop-in centers (access to condoms and weekly mobile testing on site).

\section{Methods}
\subsection{Participants}

\paragraph{Research Design, Sampling, and Recruitment.} All study procedures were approved by the University of Southern California Institutional Review Board. This is a quasi-experimental, three-group (AI, DC, OBS), pretest/posttest design, drawn from nine unique networks of YEH. A total of 714 youth were recruited across nine networks of YEH (ages 16-24) from three different drop-in centers in Los Angeles, CA. The first network was recruited in in September 2016 and the final network was recruited in October 2018. At any drop-in center, recruitment was separated by a 6-month interval to allow for sufficient numbers of new youths to replace prior clients.  

Specifics of PCA selection follow the description of the intervention delivery model. All youths receiving services were eligible to participate and were informed of the study as they entered the drop-in center. One lead study staff recruited participants throughout the study to ensure that participants were included only once. The small number of youths who accessed the drop-in center on repeat recruitment periods were only included in the sample the first time they were encountered.

\subsection{Intervention Design and Delivery} 

The details of our intervention development process have been previously published (Rice, Tulbert, et al., 2012; Rice et al., 2018). Our intervention design was based on previous literature, community collaborations, youth input, and research team members’ long-term experience working with YEH in both research and service delivery contexts. The intervention design was also informed by multiple theories, including the risk amplification and abatement model (Milburn et al., 2009) and the diffusion of innovation (Rogers, 2003). This intervention was specifically developed for YEH who access drop-in service centers. 

In both AI and DC, PCAs were recruited over a 3-week period. Each training session was limited to a maximum of five participants. To achieve the desired total number of PCAs in the network, researchers conducted three subsequent weekly trainings with a small group of participants. Training was delivered by two or three facilitators--a MSW staff lead and/or MSW student interns. The primary intervention training lasted approximately 4 hours (one half-day). Training was interactive and broken into six 45-minute modules on the mission of PCAs, sexual health, HIV prevention, communication skills, leadership skills, and self-care. PCAs were asked to focus their communications on their social ties, particularly other youths at the drop-in center, and to promote regular HIV testing and condom use. PCAs were encouraged to focus on face-to-face interactions if possible but to use social media as well. The training minimized lecture-based learning and was designed to be engaging by including a variety of learning activities: group discussion, games, journaling and reflection, experiential learning, and role-playing. The small-group setting was crucial for maintaining a safe and manageable space for youths to learn and reflect. Further, the training was developed to be an empowering experience for PCAs. Consistent language was used throughout to reiterate PCAs’ role as leaders within their community.
	
The initial training was supported by 7 weeks of 30-minute follow-up check-in sessions, which focused on positive reinforcement of PCAs’ successes in engaging peers in HIV prevention conversations, problem-solving strategies to improve future conversations, and setting goals for the week with respect to peer-to-peer conversations about HIV prevention. Because of the difficulty in scheduling and the transience of this population, the check-in schedule was flexible. PCAs could check in individually with the facilitator via phone or text. Ninety-one percent of PCAs checked in at least once; modal attendance was four sessions. PCAs received \$60 for the training and \$20 for each check-in session.

\subsection{PCA Selection Methods}

Regardless of PCA selection method, social network data were used as the basis for selecting participants. Social network data was collected from youth at baseline by interviewers. After consenting to participate in the research, youth were asked by research staff to “provide the names of other youth who you talk to regularly who use this drop-in center.” Staff explained to YEH that this list of names would be used as the input data for the algorithm that selected PCAs. Names of youth who were not study participants were discarded; the remaining names generated the “edge list” of connections among YEH in the sampled networks needed to select PCAs.

In the AI arm, we used the CHANGE algorithm to select 20\% of youth in the network to be trained as PCAs. Papers detailing the development and computational experiments of this AI selection procedure are available (Wilder, Onasch-Vera, et al., 2018). In the DC arm, we used popularity as defined by degree centrality to select PCAs—that is, the 20\% of youths who have the greatest number of ties to other youths were recruited (Valente \& Pumpuang, 2007). This is the most straightforward operationalization of selecting the “most popular” youths, which is the recommended implementation strategy in the PCA literature (Kelly et al., 1997; Valente \& Pumpuang, 2007). 

\subsection{Data Collection}

Immediately after providing informed consent, participants completed a self-administered survey assessing their demographics, sexual behaviors, HIV knowledge, and HIV testing behaviors, as well as their social networks. Three waves of survey data were collected: baseline, and follow-ups at 1 and 3 months later. A flow diagram depicts the overall study design and participant retention for each arm (see Figure 1). 
	
\subsection{Measures}

\paragraph{Demographics and Covariates.} Age, birth sex, gender identity, race/ethnicity, and sexual orientation were assessed at baseline. Participants whose gender identity differed from their assigned sex at birth, or who reported gender identity as “trans-male,” “trans-female,” “gender queer/non-conforming,” or something else, were considered transgender. Current housing situation was assessed by asking participants to select from a list of choices where they spent most of their nights in the past two weeks. Responses were used to create two dummy variables: a) shelter/transitional living program (reference: all other living situations), and b) unstably housed, which included spending most nights in an apartment, hotel, foster/group home, institution or dormitory (reference: all other living situations). When both dummy variables are included in a multivariate model, the reference group are youth who are “street/unsheltered” (i.e., spending most nights unsheltered or in a structure not fit for human habitation (e.g., abandoned building, on the street, public transit, car or RV, etc.). Relationship status (currently in a romantic/intimate relationship with a committed partner, boyfriend, girlfriend) was also assessed. 

\paragraph{Sexual Risk Behavior.} Sexual risk behavior was operationalized using two variables:  condomless anal sex (CAS) and condomless vaginal sex (CVS). First, participants were asked to list their five most recent sexual partners in the past month. For each partner, participants were asked whether they had a) vaginal sex without a condom, b) anal sex without a condom, or c) both anal and vaginal sex without a condom at their last sexual encounter. A binary variable was created indicating whether a participant had CAS (yes=1; no=0) or CVS (yes=1; no=0) with at least one sexual partner in the past month. Participants who reported no condomless sex or no sexual partners in the past month were coded as 0. For the first wave of DC participants (n = 54), CAS was assessed by asking, “In the past month, how often did you use a condom when you had anal sex?” Response options ranged from 1= “Never (0\% of the time)” to 6= “Almost always (91-100\% of the time),” and 7= “I have not had anal sex in the past month.” To create our binary CAS variable, participants were coded as 1 if they responded from 1 “Never” to 5 “Most of the time,” or coded as 0 if they responded 6 “Almost always” or “I have not had anal sex in the past month.” A similar coding scheme was employed to calculate a binary CVS variable among the first wave of DC participants . 

\paragraph{HIV Testing.} HIV testing in the past six months was assessed by asking, “When was the last time you had an HIV test?” Response options ranged from 1= “Within the last month” to 6= “I have never been tested for HIV/AIDS.” A binary variable was created such that participants who reported their last test was within the past six months were coded as 1; participants who reported their last test was more than six months or never were coded as 0. 

\paragraph{HIV Knowledge.} Participants completed a brief, six-question HIV knowledge quiz. Questions, developed by our research team, were in a true-or-false and multiple choice format, to test knowledge of HIV transmission (e.g., “You can get HIV from saliva [sharing a blunt, sharing a glass, kissing]”), HIV testing (“During the first three months after infection, it is possible for a person to still test negative for HIV”), and local, population-relevant statistics (e.g., “How many homeless youth in Los Angeles are infected with HIV?”). Percentage of correct responses was used as the main outcome measure.  

\paragraph{Social Network.} Assessment of whole (sociocentric) networks followed an event-based approach (Freeman \& Webster, 1994), wherein each network was composed of relational ties between youth receiving services within the defined boundaries of each of the three drop-in center at a given point in time. Assessment of sociometric networks was carried out by conducting an in-person interview with each participant, asking, “Who are your friends at this agency?” Participants could list up to 10 friends (alters in social network terminology). A consistent team of two research assistants who were involved in recruiting research participants at each agency then determined whether alters were also enrolled as research participants in the study. Given that network data collection was staggered for the three arms at three separate drop-in centers, this resulted in nine sociocentric networks. Sociocentric network ties at baseline were used to select peer leaders (discussed below). 

\subsection{Data Analysis}

First, descriptive statistics for key variables were calculated. Significant differences on these characteristics between arms at baseline were tested using Chi square (for binary and nominal variables) or ANOVAs (for continuous variables). Generalized Estimating Equations (GEEs), a population-averaged extension of generalized linear models for repeated-measures data (Zeger et al., 1988), were used to test intervention effects for five outcome variables: 1) CAS, 2) CVS, 3) HIV testing, and 4) HIV knowledge. An intent-to-treat model was used, whereby all participants were included in the analysis regardless of whether they reported communicating with a peer leader. GEEs predicting binary outcomes used a logit link function; and a linear GEE model was specified for HIV knowledge (a continuous variable). We specified an unstructured working correlations matrix for all GEE models, given that differences in Quasi-likelihood information criteria were negligible across different specifications. 
	
Model building progressed in two steps. First, all GEE models included terms for AI (ref: OBS), DC (ref: OBS), time, and time-by-group interaction terms to assess whether change in outcomes differed for groups over time. Each model adjusted for demographic covariates shown to be significantly correlated with the outcome variable at baseline, and an interaction between male and lesbian, gay, bisexual, or questioning (LGBQ) was added, as it improved model fit statistics (Quasi-likelihood under Independence Model Criterion). Secondly, a binary term indicating whether a participant was a PCA was added, along with a PCA-by-time interaction term, to determine whether outcomes changed over time for PCAs, relative to non-PCAs. With PCA-by-time included in the models, AI-by-time and DC-by-time interaction terms that remained significant would indicate that change in the outcomes occurred that were not due to changes in behavior of the PCAs alone. 

\subsection{Missing Data}

Some data are missing because survey items were not asked, or were asked differently, in the first wave of DC arm of study. Variables not assessed among these participants were relationship status and recent housing status. Of the 713 participants who completed the baseline survey, 245 (34\%) missed the 1-month follow-up, 300 (42\%) missed the 3-month follow-up, and 180 (25\%) missed both follow-ups. Bivariate logistic regression models indicted that participants who had greater odds of missingness at both follow-ups were mixed race ($p = .05$), had lower HIV knowledge ($p = .001$), and were more likely to be in the DC group (compared to the observation only group) at baseline. Participants with lower odds of missingness at both follow-ups were those residing in a shelter or transitional living placement ($p = .01$), and who had received services at the drop-in center more frequently ($p = .002$) and for longer periods of time ($p = .01$).

\section{Results}

\subsection{Preliminary Analyses}

Descriptive statistics for demographics and other key variables at baseline are presented in Table 1. Table 2 provides a detailed breakdown of the outcome measures under investigation. These are the unadjusted frequency distributions.  At baseline, a number of significant differences in participant characteristics were found across arms. Namely, birth sex, LGBQ identity, living situation, which drop-in center participants were recruited from (all p < .05). Therefore, these and other characteristics were included as covariates in multi-variable GEE models. 

\subsection{Outcome Analyses}

Two GEE models were run. The first was limited to intervention arms and intervention-by-time interactions. These findings are not shown in Table 3. A second GEE model, which included PCA and PCA-by-time interactions, are displayed in Table 3. 

\paragraph{Condomless Anal Sex (CAS).} The first step of the GEE model predicting CAS showed that, overall, after adjusting for covariates, CAS did not significantly change over time (OR = 1.05, 95\% CI: 0.82, 1.33). A significant group-by-time interaction was found for the AI group (OR = 0.69, 95\% CI: 0.49, 0.98), but not the DC group (OR = 0.80, 95\% CI: 0.55, 1.17). In the second step of the model, the PCA-by-time interaction was not significant, although the AI-by-time interaction remained significant (OR = 0.67, 95\% CI: 0.46, 0.95).  Figure 2 shows graphically the differences between study arms with respect to CAS.

\paragraph{Condomless Vaginal Sex (CVS).} The first step of the GEE model predicting CVS indicated that overall, CVS did not significantly change over time (OR = 0.87 95\% CI: 0.71, 1.07). There was a marginally significant AI-by-time interaction (OR = 0.78, 95\% CI: 0.57, 1.05); the DC-by-time interaction was non-significant. The PCA-by-time interaction in the second step of the model was non-significant, and the AI-by-time interaction remained marginally significant (OR = 0.75, 95\% CI: 0.55, 1.03, $p<.10$). 

\paragraph{HIV Testing.} The first step of this GEE model indicated that overall, rates of HIV testing increased over time (OR = 1.34, 95\% CI: 1.05, 1.71). However, HIV testing in the AI (OR = 0.81, 95\% CI: 0.58, 1.12) or DC groups (OR = 0.99, 95\% CI: 0.64, 1.53) did not change significantly, relative to the observation only group. In the second step of this model, a PCA-by-time term was significant (OR = 1.82, 95\% CI: 1.07, 3.09), indicating that PCAs had higher odds of being tested for HIV over time. Also, the AI-by-time term became marginally significant, but in the opposite direction (OR = 0.73, 95\% CI: 0.52, 1.03). 

\paragraph{HIV Knowledge.} Results of the first step of the GEE predicting HIV knowledge showed that overall, HIV knowledge did not significantly change over time. However, HIV knowledge significantly increased in the AI ($b = 0.04, SE = 0.01, p = 0.001$) and DC ($b = 0.04, SE = 0.01, p = 0.01$) groups, relative to the observation only group. In the second step, the PCA-by-time interaction term was significant ($b = 0.11, SE = 0.02, p < .001$), indicating that PCAs increased their HIV knowledge over time. However, the AI-by-time and DC-by-time terms were no longer significant, suggesting improvements in HIV knowledge were largely driven by knowledge increases among the PCAs. 

\section{Discussion}

There are an alarming number of YEH in the United States (e.g., Morton et al., 2018). This population continues to be a group at high risk for HIV infection (Caccamo et al., 2017; Pfeifer \& Oliver, 1997), yet there are relatively few evidence-based interventions to address this critical public health need (Arnold \& Rotheram-Borus, 2009). Given the importance of peers and social influence processes that surround sexual risk-taking (e.g., Green et al., 2013; Rice, 2010; Rice, Barman-Adhikari et al., 2012; Valente \& Auerswald, 2013), a PCA model for HIV prevention targeting YEH seems to be a promising approach (Arnold \& Rotheram-Borus, 2009; Rice, Barman-Adhikari et al., 2012). PCA approaches, however, are not always successful and who the PCAs are may be as important as the messages about behavior change they convey (Rice et al., 2018; Schneider et al., 2015). Recent pilot work suggests that this may indeed be a viable approach for YEH (Rice et al., 2018). The current study aimed to test the efficacy of a new PCA intervention for YEH, which was augmented by an AI peer selection algorithm developed by computer science collaborators. This approach aims to maximize the efficiency of message dissemination in YEH social networks. Our study utilized a quasi-experimental design with three arms: an observation only arm, a typical PCA selection method arm (i.e., namely popularity--highest degree centrality), and a PCA selection arm using the AI algorithm. Results indicate using the AI algorithm is an efficacious intervention approach. 

First, and most important, there was a significant reduction in condomless anal sex for those in the AI-arm, as indicated by the significant arm by time interaction. In the AI arm, there was a significant, 33\% reduction in the odds of condomless anal sex over time compared to the observation only arm. Whereas there were no statistically significant changes over time in the DC arm. This suggests that changes in behaviors benefit not only from a PCA intervention model designed specifically for YEH, but the AI algorithm does a better job in selecting PCAs than DC (i.e., popularity).  

Second, we observed improvements in both HIV testing and HIV knowledge over time among the subset of youth who were trained as PCAs. It is not surprising to see some HIV prevention behaviors change significantly for the PCAs but not the population as a whole.  The 20\% of youth who were identified as PCAs received 8 weeks of regular interaction with the research team, who encouraged the PCAs to learn about HIV prevention which resulted in a significant increase in HIV prevention knowledge. These youth were also regularly encouraged to engage in HIV prevention behaviors such as HIV testing. It is worth noting that while there is a significant PCA by time interaction effect, we also observed a significant change over time across all study arms in HIV testing. This finding suggests that even surveying youth about HIV testing may encourage increased HIV testing, at least in service settings where HIV testing is easily available. 

The findings of this work suggest that AI may be a useful tool to augment public health and social work intervention design. Creating and testing new behavioral health interventions with human subjects is a time-consuming and costly endeavor. Computer science relies heavily on computational experiments to test and refine algorithms. This activity allows one to discard suboptimal solutions at relatively low cost. In our case, several alternatives were tested and discarded in early computational experiments (Wilder et al., 2017; Wilder et al., 2018; Yadav et al., 2015; Yadav et al., 2016) without ever reaching the stage of field tests with human subjects.  Moreover, AI allows one to improve the efficiency of intervention processes. In our case, by formulating our approach as an influence maximization problem, we were able to create selection procedures that mathematically optimize the potential impact of PCAs selected to be trained to disseminate messages of behavior change in their networks. 

There are some limitations to the current study that must be acknowledged. First, this is a quasi-experimental design, not a randomized control trial. Because the intervention seeks to train PCAs to disseminate information within their networks, randomization at the level of individual study participants is not possible. Second, all behavioral data come from self-reports; YEH may underreport risk taking (i.e., social desirability bias), but this ought to be distributed consistently across study arms. However, because of our lack of ability to use randomization, we cannot guarantee such a distribution. Third, YEH are a very transient group and retention in the study over time is very challenging. Our study retention rate was somewhat lower than what has been reported in other studies involving longitudinal follow-up of youth experiencing homelessness (Milburn et al., 2012; Rotheram-Borus et al., 2003). However, this past research was conducted within slightly more stable populations of new runaways, most all of whom returned home (Milburn et al., 2012), or youths recruited from shelter services (Rotheram-Borus et al., 2003). Youths with chronic experiences of homelessness who are attending a drop-in center are far more difficult to track and retain. Indeed, Bender and colleagues (2016) worked with a similar drop-in-based population and had similar follow-up rates. Despite these limitations, however, the study shows highly promising results with respect to both knowledge acquisition and sexual risk behavior reduction.

\section{Conclusions and Future Directions}

Peers are critical to many behavioral health issues and can augment prevention efforts. We believe that AI can help interventionists to design increasingly impactful network-based interventions. It is important to recognize that such work need not be limited to HIV prevention efforts. Indeed, our team has been exploring how AI modelling efforts could improve suicide prevention interventions for college students (Rahmattalabi et al., 2019). Moreover, we have recently pilot tested an AI-augmented violence prevention intervention for YEH (Srivastava et al., 2019). In the past few years, we have observed a growing interest in how machine learning techniques in AI may become incorporated into health and behavioral health contexts (e.g., Ahn et al., 2016; Char et al., 2018; Verghese et al., 2018). We see the current study, however, as showing how AI and social science can collaborate beyond the sphere of predictive analytics, moving into a partnership around enhanced behavioral health and prevention intervention design.

\section{References}

\begin{itemize}[label={},itemindent=-2em,leftmargin=2em]

\item Ahn, W. Y., Ramesh, D., Moeller, F. G., \& Vassileva, J. (2016). Utility of machine-learning approaches to identify behavioral markers for substance use disorders: Impulsivity dimensions as predictors of current cocaine dependence. Frontiers in Psychiatry, 7, 34. doi: 10.3389/fpsyt.2016.00034

\item Amirkhanian, Y. A., Kelly, J. A., Kabakchieva, E., McAuliffe, T. L., \& Vassileva, S. (2003). Evaluation of a social network HIV prevention intervention program for young men who have sex with men in Russia and Bulgaria. AIDS Education and Prevention, 15(3), 205–220. doi:10 .1521/aeap.15.4.205.23832

\item Arnold, E. M., \& Rotheram-Borus, M. J. (2009). Comparisons of prevention programs for homeless youth. Prevention Science, 10(1), 76–86. \url{https://doi.org/10.1007/s11121-008-0119-4}

\item Bender, K. A., DePrince, A., Begun, S., Hathaway, J., Haffejee, B., \& Schau, N. (2016). Enhancing risk detection among homeless youth: A randomized clinical trial of a promising pilot intervention. Journal of Interpersonal Violence, 33(19) 2945–2967. doi:107.171/0778 /806882620650156166333208 

\item Caccamo, A., Kachur, R., \& Williams, S. P. (2017). Narrative review: Sexually transmitted diseases and homeless youth—What do we know about sexually transmitted disease prevalence and risk?. Sexually Transmitted Diseases, 44(8), 466--476. doi:10.1097/OLQ.0000000000000633

\item Char, D. S., Shah, N. H., \& Magnus, D. (2018). Implementing machine learning in health care—Addressing ethical challenges. The New England Journal of Medicine, 378(11), 981--983. doi: 10.1056/NEJMp1714229

\item Freeman, L. C., \& Webster, C. M. (1994). Interpersonal proximity in social and cognitive space. Social Cognition, 12(3), 223–247. \url{https://doi.org/10.1521/soco.1994.12.3.223}

\item Green  Jr., H. D., de la Haye, K., Tucker, J. S., \& Golinelli, D. (2013). Shared risk: Who engages in substance use with American homeless youth? Addiction, 108(9), 1618–1624. \url{https://doi.org/http://dx.doi.org/10.1111/add.12177}

\item Kelly, J. A., Murphy, D. A., Sikkema, K. J., McAuliffe, T. L., Roffman, R. A., Solomon, L. J., Winett, R. A., \& Kalichman, S. C. (1997). Randomised, controlled, community-level HIV-prevention intervention for sexual-risk behaviour among homosexual men in US cities. Lancet, 350(9090), 1500–1505. \url{https://doi.org/10.1016/S0140-6736(97)07439-4}

\item Kelly, J. A. (2004). Popular opinion leaders and HIV prevention peer education: Resolving discrepant findings, and implications for the development of effective community programmes. AIDS Care, 16(2), 139–150. doi:10.1080/09540120410001640986

\item Kempe, D., Kleinberg, J., \& Tardos, E. (2003). Maximizing the spread of influence through a social network. Proceedings of the Ninth ACM SIGKDD International Conference on Knowledge Discovery and Data Mining (pp. 137–146). Retrieved from \url{http://www.theoryofcomputing.org /articles/v011a004/v011a004.pdf}

\item Latkin, C. A., Sherman, S., \& Knowlton, A. (2003). HIV prevention among drug users: Outcome of a network-oriented peer outreach intervention. Health Psychology, 22(4), 332–339. \url{https://doi.org/10.1037/0278-6133.22.4.332}

\item Medley, A., Kennedy, C., O’Reilly, K., \& Sweat, M. (2009). Effectiveness of peer education interventions for HIV prevention in developing countries: A systematic review and meta-analysis. AIDS Education and Prevention, 21(3), 181–206. \url{https://doi.org/10.1521/aeap.2009.21.3.181}

\item Milburn, N. G., Rice, E., Rotheram-Borus, M. J., Mallett, S., Rosenthal, D., Batterham, P., May, S. J., Witkin, A., \& Duan, N. (2009). Adolescents exiting homelessness over two years: The Risk Amplification and Abatement Model. Journal of Research on Adolescence, 19(4), 762–785. \url{https://doi.org/10.1111/j.1532-7795.2009.00610.x}

\item Milburn, N. G., Iribarren, F. J., Rice, E., Lightfoot, M., Solorio, R., Rotheram-Borus, M. J., Desmond, K., Lee, A., Alexander, K., Maresca, K., Eastmen, K., Arnold, E. M., \& Duan, N. (2012). A family intervention to reduce sexual risk behavior, substance use, and delinquency among newly homeless youth. Journal of Adolescent Health, 50(4), 358– 364. doi:10.1016/j.jadohealth.2011.08.009 

\item Morton, M. H., Dworsky, A., Matjasko, J. L., Curry, S. R., Schlueter, D., Chávez, R., \& Farrell, A. F. (2018). Prevalence and correlates of youth homelessness in the United States. Journal of Adolescent Health, 62(1), 14–21. \url{https://doi.org/10.1016/j.jadohealth.2017.10.006}

\item NIMH Collaborative HIV/STD Prevention Trial Group. (2010). Results of the NIMH collabora- tive HIV/STD prevention trial of a community popular opinion leader intervention. Journal of Acquired Immune Deficiency Syndromes, 54(2), 204–214. doi:10.1097/QAI.0b013e3181d61def 

\item Pfeifer, R. W., \& Oliver, J. (1997). A study of HIV seroprevalence in a group of homeless youth in Hollywood, California. Journal of Adolescent Health, 20(5), 339–342. \url{https://doi.org/10.1016/S1054-139X(97)00038-4}

\item Rahmattalabi, A., Vayanos, P., Fulginiti, A., \& Tambe, M. (2019, May). Robust peer-monitoring on graphs with an application to suicide prevention in social networks. Proceedings of the 18th International Conference on Autonomous Agents and MultiAgent Systems (pp. 2168-2170).

\item Rice, E. (2010). The positive role of social networks and social networking technology in the condom-using behaviors of homeless young people. Public Health Reports, 125(4), 588–595. \url{https://doi.org/10.1177/003335491012500414}

\item Rice, E., Barman-Adhikari, A., Milburn, N. G., \& Monro, W. (2012). Position-specific HIV risk in a large network of homeless youths. American Journal of Public Health, 102(1), 141–147. \url{https://doi.org/10.2105/AJPH.2011.300295}

\item Rice, E., Tulbert, E., Cederbaum, J., Barman Adhikari, A., \& Milburn, N. G. (2012). Mobilizing homeless youth for HIV prevention: A social network analysis of the acceptability of a face-to-face and online social networking intervention. Health Education Research, 27(2), 226--236. doi: 10.1093/her/cyr113

\item Rice, E., Yoshioka-Maxwell, A., Petering, R., Onasch-Vera, L., Craddock, J., Tambe, M., Yadav, A., Wilder, B., Woo, D., Winetrobe, H., \& Wilson, N. (2018). Piloting the use of artificial intelligence to enhance HIV prevention interventions for youth experiencing homelessness. Journal of the Society for Social Work and Research, 9(4), 551–573. \url{https://doi.org/10.1086/701439}

\item Rogers, E. M. (2003). Diffusion of Innovations (5th ed.). New York, NY: Free Press.

\item Rotheram-Borus, M. J., Song, J., Gwadz, M., Lee, M., Van Rossem, R., \& Koopman, C. (2003). Reductions in HIV risk among runaway youth. Prevention Science, 4(3), 173–187. \url{https://doi.org/10.1023/A:1024697706033}

\item Schneider, J. A., Zhou, A. N., \& Laumann, E. O. (2015). A new HIV prevention network approach: Sociometric peer change agent selection. Social Science \& Medicine, 125, 192–202. \url{https://doi.org/10.1016/j.socscimed.2013.12.034}

\item Slesnick, N., Feng, X., Guo, X., Brakenhoff, B., Carmona, J., Murnan, A., Cash, S., \& McRee, A.-L. (2016). A test of outreach and drop-in linkage versus shelter linkage for connecting homeless youth to services. Prevention Science, 17(4), 450–460. \url{https://doi.org/10.1007/s11121-015-0630-3}

\item Srivastava, A., Petering, R., Barr, N., Kannan, R., Rice, E., \& Prasanna, V. K. (2019). Network-based intervention strategies to reduce violence among homeless. Social Network Analysis and Mining, 9, 38. \url{https://doi.org/10.1007/s13278-019-0584-8}

\item Valente, A. M., \& Auerswald, C. L. (2013). Gender differences in sexual risk and sexually transmitted infections correlate with gender differences in social networks among San Francisco homeless youth. The Journal of Adolescent Health, 53(4), 486–491. \url{https://doi.org/10.1016/j.jadohealth.2013.05.016}

\item Valente, T. W., \& Pumpuang, P. (2007). Identifying opinion leaders to promote behavior change. Health Education and Behavior, 34(6), 881–896. \url{https://doi.org/10.1177/1090198106297855}

\item Verghese, A., Shah, N. H., \& Harrington, R. A. (2018). What this computer needs is a physician: humanism and artificial intelligence. Journal of the American Medical Association, 319(1), 19--20. doi:10.1001/ jama.2017.19198

\item Wilder, B., Yadav, A., Immorlica, N., Rice, E., \& Tambe, M. (2017, May). Uncharted but not uninfluenced: Influence maximization with an uncertain network. Proceedings of the 16th Conference on Autonomous Agents and MultiAgent Systems, pp. 1305--1313.

\item Wilder, B., Immorlica, N., Rice, E., \& Tambe, M. (2018, April). Maximizing influence in an unknown social network. Proceedings of the 32nd AAAI Conference on Artificial Intelligence, pp. 4743--4750.

\item Wilder, B., Onasch-Vera, L., Hudson, J., Luna, J., Wilson, N., Petering, R., Woo, D., Tambe, M., \& Rice, E. (2018, July). End-to-end influence maximization in the field. Proceedings of the 17th International Conference on Autonomous Agents and MultiAgent Systems, pp. 1414--1422.

\item Yadav, A., Chan, H., Xin Jiang, A., Xu, H., Rice, E., \& Tambe, M. (2016) Using social networks to aid homeless shelters: Dynamic influence maximization under uncertainty. Proceedings of the 15th International Conference on Autonomous Agents and Multiagent Systems, pp. 740–748. 

\item Yadav, A., Marcolino, L. S., Rice, E., Petering, R., Winetrobe, H., Rhoades, H., Tambe, M., \& Carmichael, H. (2015). Preventing HIV spread in homeless populations using PSINET. Proceedings of the 27th Conference on Innovative Applications of Artificial Intelligence, pp. 4006--4011. 

\item Yadav, A., Wilder, B., Rice, E., Petering, R., Craddock, J., Yoshioka-Maxwell, A., Hemler, M., Onasch-Vera, L., Tambe, M., \& Woo, D. (2018, July). Bridging the gap between theory and practice in influence maximization: Raising awareness about HIV among homeless youth. Proceedings of the 27th International Joint Conference on Artificial Intelligence pp. 5399--5403.

\item Yadav, A., Wilder, B., Rice, E., Petering, R., Craddock, J., Yoshioka-Maxwell, A., Hemler, M., Onasch-Vera, L., Tambe, M., \& Woo, D. (2017, May). Influence maximization in the field: The arduous journey from emerging to deployed application. Proceedings of the 16th Conference on Autonomous Agents and MultiAgent Systems, pp. 150--158.

\item Zeger, S. L., Liang, K. Y., \& Albert, P. S. (1988). Models for longitudinal data: A generalized estimating equation approach. Biometrics, 44(4), 1049–1060.

\end{itemize}

\begin{table}
    \centering 
    \caption{Participant Characteristics at Baseline (N = 714).}
    \fontsize{9}{9}\selectfont
    \begin{tabular}{lcccc}
    \toprule
          & 
AI (n = 253)
 & 
DC (n = 209)
 & 
CTL (n = 251)
 & 
Full Sample 
 \\ 
        
Variable
 & 
N (\%) or M (SD)
 & 
N (\%) or M (SD)
 & 
N (\%) or M (SD)
 & 
N (\%) or M (SD)
 \\ \midrule
        
Age
 & 
22.0 (2.0)
 & 
21.8 (2.2)
 & 
21.8 (2.2)
 & 
21.9 (2.1)
 \\ 
        
\textbf{Gender}
 & 
 
 & 
 
 & 
 
 & 
 
 \\ 
        
  Male*
 & 
193 (76.3\%)
 & 
164 (78.5\%)
 & 
175 (69.7\%)
 & 
533 (74.6\%)
 \\ 
        
  Female
 & 
60 (23.7\%)
 & 
42 (20.1\%)
 & 
76 (30.3\%)
 & 
178 (24.9\%)
 \\ 
        
  Transgender/non-binary
 & 
32 (12.6\%)
 & 
25 (12.0\%)
 & 
37 (14.7\%)
 & 
94 (13.2\%)
 \\ 
        
\textbf{Race/Ethnicity}
 & 
 
 & 
 
 & 
 
 & 
 
 \\ 
        
  Black/African-American
 & 
88 (34.8\%)
 & 
64 (30.6\%)
 & 
69 (27.5\%)
 & 
221 (31.0\%)
 \\ 
        
  Non-Hispanic White
 & 
 
 & 
 
 & 
 
 & 
160 (22.4\%)
 \\ 
        
  Hispanic Latino/a/x
 & 
33 (13.0\%)
 & 
27 (12.9\%)
 & 
45 (17.9\%)
 & 
106 (14.8\%)
 \\ 
        
  Multiple
 & 
70 (27.7\%)
 & 
50 (23.9\%)
 & 
59 (23.5\%)
 & 
179 (25.1\%)
 \\ 
        
  Other$^\text{a}$
 & 
15 (5.9\%)
 & 
14 (6.7\%)
 & 
19 (7.6\%)
 & 
48 (6.7\%)
 \\ 
        
LGBQA*
 & 
120 (47.4\%)
 & 
74 (35.4\%)
 & 
112 (44.6\%)
 & 
306 (42.9\%)
 \\ 
        
Romantic Relationship (Current)
 & 
88 (34.8\%)
 & 
66 (31.6\%)$^\text{b}$
 & 
96 (38.2\%)
 & 
251 (35.2\%)
 \\ 
        
\textbf{Housing}
 & 
 
 & 
 
 & 
 
 & 
 
 \\ 
        
  Shelter/Transitional Living
 & 
57 (22.5\%)
 & 
56 (26.8\%)$^\text{b}$
 & 
47 (18.7\%)
 & 
160 (22.4\%)
 \\ 
        
  Unstably Housed*
 & 
94 (37.2\%)
 & 
99 (47.4\%)$^\text{b}$
 & 
94 (37.5\%)
 & 
287 (40.2\%)
 \\ 
        
  Street (Unsheltered)*  
 & 
102 (40.3\%)
 & 
54 (25.8\%)$^\text{b}$
 & 
110 (43.8\%)
 & 
267 (37.4\%)
 \\ 
        
\textbf{Drop-in*}
 & 
 
 & 
 
 & 
 
 & 
 
 \\ 
        
  My Friend’s Place
 & 
64 (25.3\%)
 & 
74 (35.4\%)
 & 
90 (35.9\%)
 & 
228 (31.9\%)
 \\ 
        
  Youth Center on Highland
 & 
96 (37.9\%)
 & 
81 (38.8\%)
 & 
80 (31.9\%)
 & 
257 (36.0\%)
 \\ 
        
  Safe Place for Youth
 & 
93 (36.8\%)
 & 
54 (25.8\%)
 & 
81 (32.3\%)
 & 
228 (31.9\%)
 \\ \bottomrule
    \end{tabular}

\begin{flushleft}
*Significant between-group differences at $p < .05$

$^\text{a}$Other race includes Asian, Native, Pacific Islander, and “other” as a write in category

$^\text{b}$Data missing from first wave of DC (n = 54)
\end{flushleft}
\end{table}

\clearpage

\begin{table}
    \centering
    \caption{Outcomes Over Time, Stratified by Intervention Group (N = 714).}
    \fontsize{9}{9}\selectfont
    \begin{tabular}{l|cc|cc|cc}
    \toprule
        
Variable
 & 
 \multicolumn{2}{c}{AI}
 & 
 \multicolumn{2}{c}{DC}
 & 
 \multicolumn{2}{c}{OBS}
 \\ \midrule
        
Condomless Anal Sex (CAS)
 & 
n (\%)
 & 
Total N
 & 
n (\%)
 & 
Total N
 & 
n (\%)
 & 
Total N
 \\ \midrule
        
Baseline
 & 
69 (27\%)
 & 
253
 & 
69 (33\%)
 & 
205
 & 
54 (22\%)
 & 
251
 \\ 
        
1M
 & 
31 (18\%)
 & 
171
 & 
43 (35\%)
 & 
124
 & 
37 (21\%)
 & 
176
 \\ 
        
3M
 & 
27 (18\%)
 & 
154
 & 
26 (24\%)
 & 
108
 & 
36 (24\%)
 & 
152
 \\ \midrule
        
Condomless Vaginal Sex (CVS)
 & 
n (\%)
 & 
Total N
 & 
n (\%)
 & 
Total N
 & 
n (\%)
 & 
Total N
 \\ \midrule
        
Baseline
 & 
90 (36\%)
 & 
253
 & 
87 (42\%)
 & 
206
 & 
116 (46\%)
 & 
251
 \\ 
        
1M
 & 
51 (30\%)
 & 
171
 & 
53 (43\%)
 & 
124
 & 
62 (35\%)
 & 
176
 \\ 
        
3M
 & 
31 (20\%)
 & 
154
 & 
34 (32\%)
 & 
108
 & 
61 (40\%)
 & 
152
 \\ \midrule
        
HIV testing (past six months)
 & 
n (\%)
 & 
Total N
 & 
n (\%)
 & 
Total N
 & 
n (\%)
 & 
Total N
 \\ \midrule
        
Baseline
 & 
183 (74\%)
 & 
246
 & 
139 (67\%)
 & 
207
 & 
181 (72\%)
 & 
249
 \\ 
        
1M
 & 
136 (82\%)
 & 
166
 & 
101 (80\%)
 & 
126
 & 
141 (81\%)
 & 
174
 \\ 
        
3M
 & 
114 (75\%)
 & 
152
 & 
79 (74\%)
 & 
107
 & 
123 (83\%)
 & 
149
 \\ \midrule
        
HIV Knowledge Test (\% Correct)
 & 
M (SD)
 & 
Total N
 & 
M (SD)
 & 
Total N
 & 
M (SD)
 & 
Total N
 \\ \midrule
        
Baseline
 & 
58\% (20\%)
 & 
251
 & 
56\% (22\%)
 & 
154
 & 
57\% (20\%)
 & 
249
 \\ 
        
1M
 & 
69\% (22\%)
 & 
167
 & 
64\% (23\%)
 & 
88
 & 
63\% (20\%)
 & 
174
 \\ 
        
3M
 & 
65\% (23\%)
 & 
152
 & 
63\% (25\%)
 & 
74
 & 
59\% (19\%)
 & 
150
 \\ \bottomrule
    \end{tabular}
\end{table}

\clearpage

\begin{table}
    \centering
    \caption{Including all outcomes in one table. Generalized Estimating Equations Predicting Study Outcome Variables, including PCA and PCA x Time Interaction.}
    \begin{tabular}{llcccccc}
    \toprule
        
Outcome Variable
 & 
Predictor
 & 
$b$
 & 
$SE$
 & 
 OR
 & 
95\% CI
 & 
$p$
 & 
 
 \\ \midrule
        
Condomless Anal Sex (CAS)
 & 
 
 \\ 
          & 
Time
 & 
--
 & 
--
 & 
1.05
 & 
0.82, 1.33
 & 
0.72
 & 
 
 \\ 
          & 
AI
 & 
--
 & 
--
 & 
1.44
 & 
0.89, 2.32
 & 
0.14
 & 
 
 \\ 
          & 
DC
 & 
--
 & 
--
 & 
1.49
 & 
0.88, 2.52
 & 
0.14
 & 
 
 \\ 
        
 
 & 
PCA
 & 
--
 & 
--
 & 
1.00
 & 
0.56, 1.78
 & 
0.99
 & 
 
 \\ 
          & 
\textbf{AI x Time}
 & 
--
 & 
--
 & 
\textbf{0.67
} & 
\textbf{0.46, 0.95
} & 
\textbf{0.03
} & 
 
 \\ 
          & 
DC x Time
 & 
--
 & 
--
 & 
0.78
 & 
0.52, 1.15
 & 
0.20
 & 
 
 \\ 
        
 
 & 
PCA x Time
 & 
--
 & 
--
 & 
1.17
 & 
0.73, 1.89
 & 
0.51
 & 
 
 \\ 
        
Condomless Vaginal Sex (CVS)
 & 
 
 \\ 
          & 
Time
 & 
--
 & 
--
 & 
0.87
 & 
0.71, 1.07
 & 
0.18
 & 
 
 \\ 
          & 
AI
 & 
--
 & 
--
 & 
0.79
 & 
0.53, 1.18
 & 
0.25
 & 
 
 \\ 
          & 
DC
 & 
--
 & 
--
 & 
1.09
 & 
0.68, 1.74
 & 
0.72
 & 
 
 \\ 
          & 
PCA
 & 
--
 & 
--
 & 
0.86
 & 
0.49, 1.50
 & 
0.59
 & 
 
 \\ 
          & 
\textit{AI x Time
} & 
--
 & 
--
 & 
\textit{0.75
} & 
\textit{0.55, 1.03
} & 
\textit{0.08
} & 
 
 \\ 
        
 
 & 
DC x Time
 & 
--
 & 
--
 & 
0.85
 & 
0.60, 1.22
 & 
0.38
 & 
 
 \\ 
        
 
 & 
PCA x Time
 & 
--
 & 
--
 & 
1.17
 & 
0.76, 1.81
 & 
0.47
 & 
 
 \\ 
        
HIV Testing
 & 
 
 \\ 
          & 
\textbf{Time}
 & 
--
 & 
--
 & 
\textbf{1.34
} & 
\textbf{1.05, 1.71
} & 
\textbf{0.02
} & 
 
 \\ 
          & 
AI
 & 
--
 & 
--
 & 
1.21
 & 
0.77, 1.91
 & 
0.40
 & 
 
 \\ 
          & 
DC
 & 
--
 & 
--
 & 
0.95
 & 
0.58, 1.56
 & 
0.84
 & 
 
 \\ 
          & 
PCA
 & 
--
 & 
--
 & 
0.79
 & 
0.41, 1.53
 & 
0.48
 & 
 
 \\ 
        
 
 & 
\textit{AI x Time
} & 
--
 & 
--
 & 
\textit{0.73
} & 
\textit{0.52, 1.03
} & 
\textit{0.07
} & 
 
 \\ 
        
 
 & 
DC x Time
 & 
--
 & 
--
 & 
0.91
 & 
0.58, 1.41
 & 
0.66
 & 
 
 \\ 
          & 
\textbf{PCA x Time
} & 
--
 & 
--
 & 
\textbf{1.82
} & 
\textbf{1.07, 3.09
} & 
\textbf{0.03
} & 
 
 \\ 
        
HIV Knowledge
 \\ 
          & 
Time
 & 
0.002
 & 
0.009
 & 
--
 & 
-0.02, 0.02
 & 
0.79
 & 
 
 \\ 
          & 
AI
 & 
-0.02
 & 
0.02
 & 
--
 & 
-0.05, 0.02
 & 
0.34
 & 
 
 \\ 
          & 
\textit{DC}
 & 
\textit{-0.04
} & 
\textit{0.02
} & 
--
 & 
\textit{-0.08, 0.01
} & 
\textit{0.09
} & 
 
 \\ 
          & 
PCA
 & 
0.02
 & 
0.03
 & 
--
 & 
-0.03, 0.07
 & 
0.38
 & 
 
 \\ 
          & 
AI x Time
 & 
0.02
 & 
0.01
 & 
--
 & 
-0.01, 0.04
 & 
0.20
 & 
 
 \\ 
        
 
 & 
DC x Time
 & 
0.02
 & 
0.02
 & 
--
 & 
-0.01, 0.05
 & 
0.27
 & 
 
 \\ 
        
 
 & 
\textbf{PCA x Time
} & 
\textbf{0.11
} & 
\textbf{0.02
} & 
--
 & 
\textbf{0.07, 0.16
} & 
\textbf{$<$ 0.001
} & 
 
 \\ \bottomrule
    \end{tabular}
    
    \begin{flushleft}
    \textit{Note}: Models adjusted for age, male birth sex, transgender, LGBQ identity, male X LGBQ, race, committed relationship, housing status, drop-in center. 
    
\textit{Note}: \textbf{Bolded} parameter estimates are significant at $p < .05$.

\textit{Note}: \textit{Italicized} parameter estimates are significant at $p < .10$.

    \end{flushleft}

\end{table}

\clearpage

\begin{figure}
    \centering
    \includegraphics[width=6in]{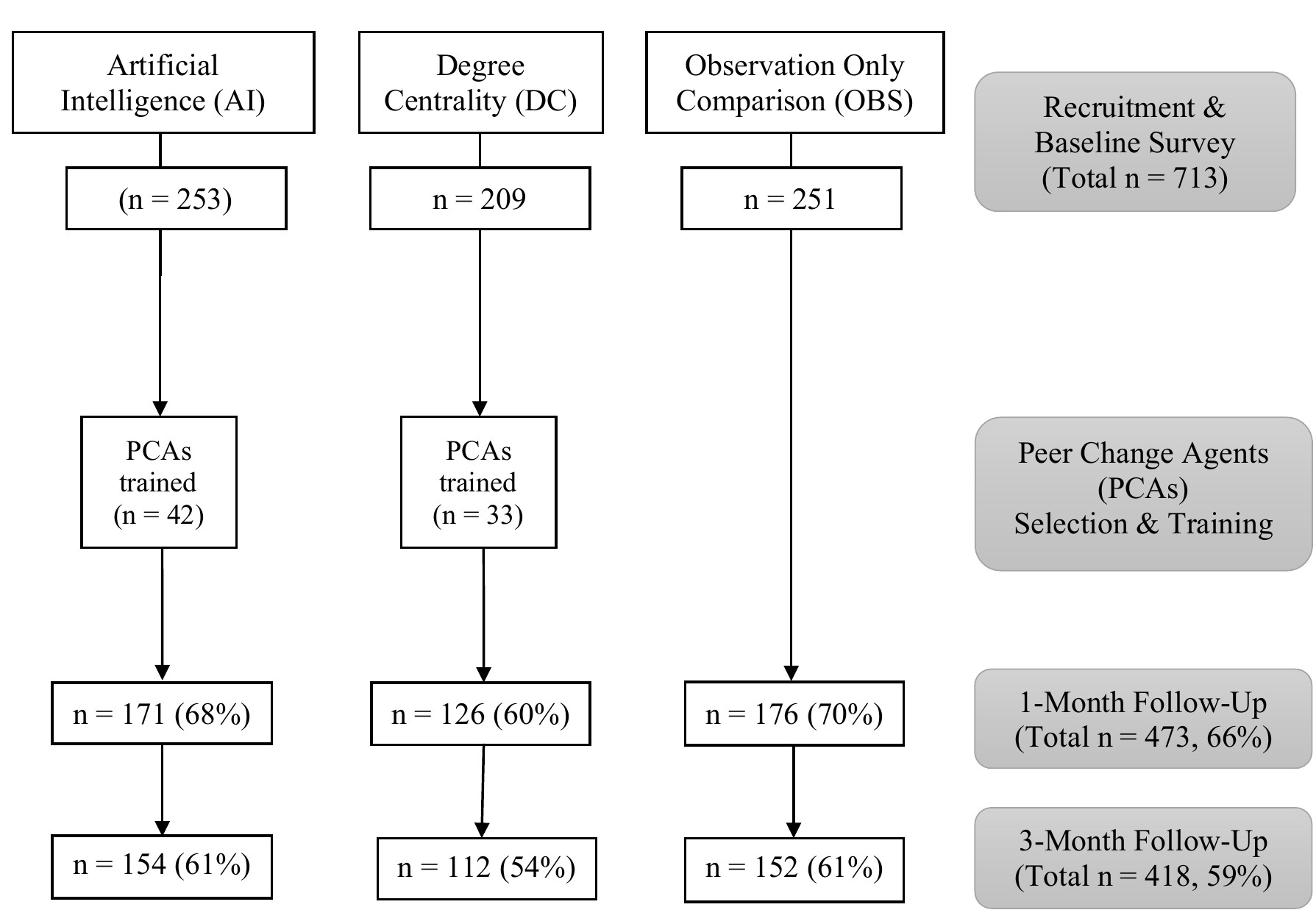}
    \caption{Participant Flow Diagram.}
    \label{fig:participants}
\end{figure}

\clearpage

\begin{figure}
    \centering
    \includegraphics[width=4in]{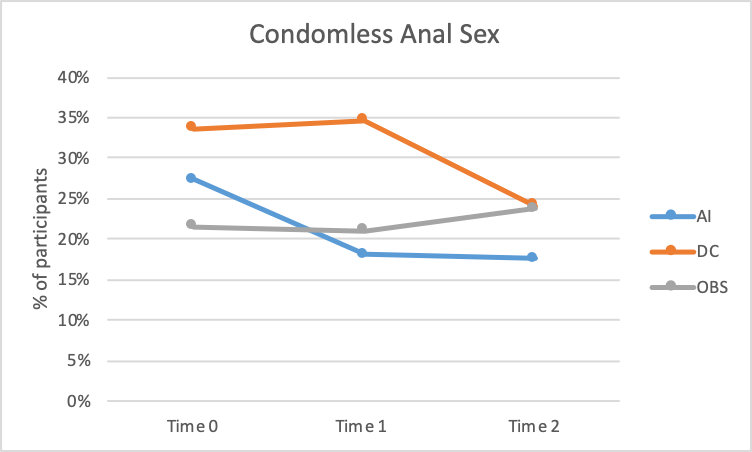}
    \caption{Prevalence of Condomless Anal Sex by Study Arm. AI = Artificial Intelligence. DC = Degree Centrality. OBS = Observation only control.}
    \label{fig:cas}
\end{figure}

\end{document}